# Dark Field X-ray Microscopy Below Liquid-Helium Temperature: The Case of NaMnO$_2$

Jayden Plumb[a,b], Ishwor Poudyal[b,c], Rebecca L. Dally[d], Samantha Daly[a], Stephen D. Wilson[e], Zahir Islam[b, *]

[a] Department of Mechanical Engineering, University of California, Santa Barbara, Santa Barbara, California 93106, United States
[b] X-ray Science Division, Argonne National Laboratory, 9700 S Cass Ave, Lemont, Illinois 60439, United States
[c] Materials Science Division, Argonne National Laboratory, 9700 S Cass Ave, Lemont, Illinois 60439, United States
[d] NIST Center for Neutron Research, National Institute of Standards and Technology, 100 Bureau Dr, Gaithersburg, MD 20899, United States
[e] Materials Department, University of California, Santa Barbara, Santa Barbara, California 93106, United States

## Abstract

Dark field X-ray microscopy (DFXM) is an experimental technique employed to investigate material properties by probing their 'mesoscale,' or microscale structures, in a bulk-sensitive manner using hard X-rays at synchrotron radiation sources. However, challenges remain when it comes to applications of this technique to examine low-temperature phenomena in quantum materials, which exhibit complex phase transitions at cryogenic temperatures. One such material is NaMnO$_2$, which hosts an antiferromagnetic transition at 45 K that is suspected to coincide with local structural transitions from its majority monoclinic phase to nanoscale triclinic domains. Direct observation of local heterogeneities and this effect at low temperatures in NaMnO$_2$ is an important step in understanding this material, and serves as an ideal candidate study for expanding the DFXM experimental design space. This paper details a foundational high-resolution DFXM study, down to liquid-helium temperature and below, conducted to explore phase transitions in NaMnO$_2$. The outlined experiment ushers in the evaluation of other functional materials at low temperatures using this technique.

* Corresponding Author
Email: zahir@anl.gov

# 1. Introduction

Dark field X-ray microscopy (DFXM) is a non-destructive bulk-sensitive technique that provides a high-accuracy spatially resolved map of an ordered material's 'mesoscale' structure [1]. These maps possess information about the local crystallographic orientation spread (such as mosaic and twins) and lattice strain heterogeneities. DFXM has been used to study structure-property relationships in a variety of materials (Refs. [2,3] and the references therein), but its effective use beyond room-temperature studies remains a challenge, particularly at the cryogenic temperatures frequently encountered in investigations of quantum materials [4]. This is due to competing requirements of low vibrations and mechanical stability, versus the operational demands and physical connections of a cryostat mounted on an X-ray diffractometer. Here, we demonstrate an advancement of this technique, leveraging a low-vibration cryostat uniquely adapted to an unobstructed horizontal-diffraction geometry, and detail a novel DFXM study conducted at cryogenic temperatures (see Figure 1). This experiment included collecting magnified-real-space diffraction-contrast images using a standard rocking curve scan methodology, with the unique benefit of *in-situ* cooling and heating. The range of temperatures achieved spanned 3.2 K to 300 K, which establishes an original approach to characterizing thermal phase transformations of crystal, electronic, and magnetic structures.

In DFXM, a real-space image is formed by placing an X-ray objective lens into the diffracted beam. The objective, commonly a compound refractive lens (CRL) well-suited for hard X-ray energies, magnifies the diffracted beam and projects it onto a far-field area detector that is placed at the image plane. This provides a full-field image due to diffraction contrasts from a plethora of heterogeneities (*e.g.*, twins, charge and/or magnetic domains, dislocations, competing phases, impurities, etc.) with a high spatial resolution on the order of 200 nanometers (transverse to a diffracted beam). A sample-to-detector distance ranging in meters enables high angular resolution as well, on the order of 0.001°. The use of an objective lens with a focal length on the order of 100s of millimeters, leaves space for a cryostat, discussed below, and other sample environments (*i.e.*, load frame, magnet, etc.).

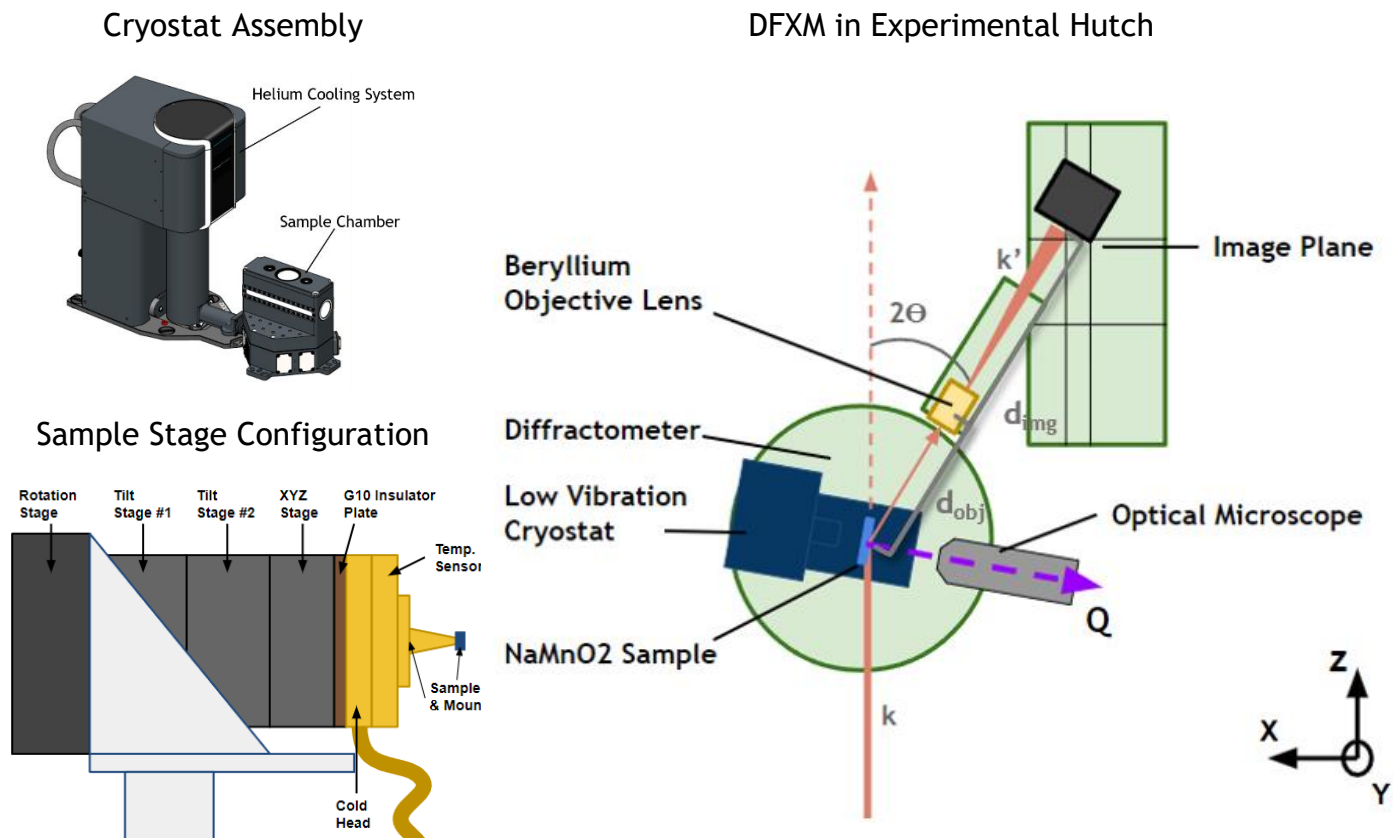

Figure 1: Schematic of the DFXM set-up (right) with a low-vibration cryostat (upper left) equipped with a cryogenic sample tower (lower left). The sample is mounted to an integrated highly efficient, custom thermal link (shown in gold). Thermal mounting stages are attached to a cryogenic sample tower consisting of tip-tilt, translation, and rotation stages with nanoscale precision.

$NaMnO_2$ has gained attention due to an ever-increasing demand for energy transformation and storage, requiring modern electronic and battery applications to exploit functional materials with exceptional properties. This field of research has recently turned toward layered sodium oxides, such as $NaMnO_2$, for use as cathodes, due to their high reversible capacities and the low cost of earth-abundant sodium. Its α-phase consists of layers of $MnO_6$ octahedra separated by layers of Na ions that form a monoclinic C2/m crystal structure. Furthermore, its room-temperature electronic properties are complemented by a rich magnetic phase space at low temperatures. The $MnO_6$ octahedra mentioned above are subject to a Jahn-Teller distortion, resulting in short-range quasi-1D antiferromagnet interactions that are eventually driven to three-dimensional order below the Néel temperature of 45 K. This antiferromagnetic state remains incommensurate with the crystal lattice until another transition to a commensurate phase occurs at 22 K [5]. The functional properties exhibited by this material are further complicated by a series of crystal defects (*i.e.*, stacking faults, twinning, polymorphism and an $Mn_3O_4$ intergrowth presented in Figure 2).

Thus, there are several sources of meso- and microscale heterogeneities in $NaMnO_2$ that can significantly impact its macroscopic behavior via local strain fields, spatial variance, and lattice rotation, making DFXM an ideal probe. DFXM has been used to identify individual dislocations and the strain fields around them [6-8]; a DFXM study of crystal defects in steel (twinning, stacking faults, precipitates, etc.), identified defect features in the collected images [9].

In $NaMnO_2$, twin populations have been correlated to its electrochemical potential [10], while the native strains induced from twinning in a similar Ni- or Co- doped sodium manganese oxide have been linked to its failure during high-voltage charge cycling [11]. Additionally, a structural transition from a monoclinic to triclinic crystal structure has been postulated to occur locally within $NaMnO_2$ at 45 K [12], but the effect is subtle and a lack of spatial resolution leaves this question largely unanswered. These examples highlight the need for high-accuracy, spatially resolved measurements to better understand hierarchical structural complexity and its impact on the functional behavior of $NaMnO_2$. In particular, the impact of the low-temperature magneto-structural transition of $NaMnO_2$ on its structural complexity is the focus of this work.

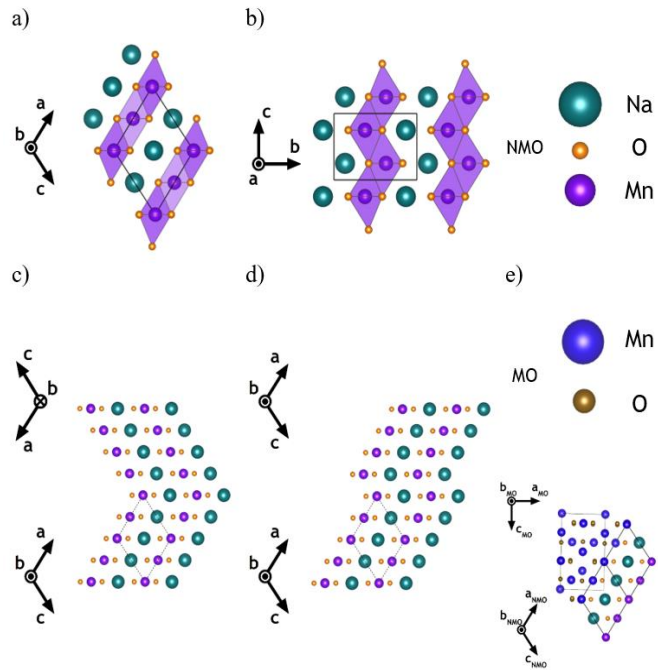

Figure 2: Crystal structure of the a) α-phase and b) β-phase of $NaMnO_2$. Examples of crystal structure defects in the α-phase, including a c) twin, d) stacking fault and e) $Mn_3O_4$ intergrowth. The various unit cells are outlined using solid or dashed black lines.

## 2. Material and Methods

### 2.1 Sample Preparation

The single-crystal $NaMnO_2$ sample was grown using the floating zone technique [13] and characterized previously using neutron scattering [5,14,15]. The high angular resolution of DFXM requires knowledge of the sample orientation prior to the experiment. As determined via Laue diffraction, $NaMnO_2$ forms a facet during growth along the (-101) lattice plane, which serves as a cleavage plane during sample preparation, with the short b-axis extending parallel to the growth direction. The sample was cleaved along this plane and cut to a size of approximately 20 mm x 3 mm x 2 mm for the room temperature survey and 6 mm x 2 mm x 0.1 mm for the low temperature experiment.

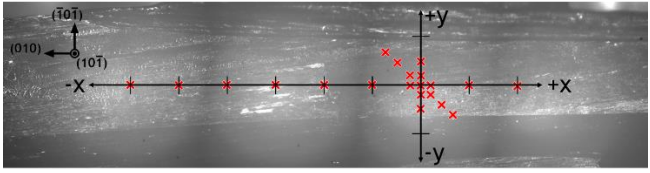

Figure 3: Series of images taken of the sample during a room-temperature survey that were stitched together using a grid/collection stitching plugin [16] and ImageJ [17]. An axis is overlaid to indicate the position at which DFXM scans were conducted over the sample area, and notches are placed at mm increments. Red Xs indicate the locations where scans were conducted. The upper left axis indicates the orientation of the single crystal sample, defined using Laue diffractometry.

Solid-state nuclear magnetic resonance (ssNMR) experiments conducted previously on samples grown using this method estimate the population of stacking faults and β-phase content with reference to the primary α-phase. The sample used here derived its Na content from approximately 96% α-phase, 4% stacking fault regions and < 1% β-phase [13]. The relatively low β-phase content of the sample enabled our DFXM study to focus primarily on the effects of twinning and stacking faults within the majority α-phase. However, the additional presence of $Mn_3O_4$ intergrowths, albeit minute, may contribute to diffraction contrast, creating difficulty in uniquely identifying some defects.

The sample was glued to a sample mount using GE varnish (see Figure 1) with its (2,0,-2) reciprocal lattice vector oriented in the horizontal scattering plane for a reflection geometry. It was then positioned and aligned with the X-ray beam center using a video microscope unit (VMU), configured with a 10x objective and a 5-megapixel Flir CMOS camera, oriented with the optical axis aligned to the crystal surface normal. A single-photon-counting sodium iodide point detector was initially used to optimize the (2,0,-2) Bragg peak (2θ=30.5° at 10 keV) and to characterize its profile aided by low-temperature cryogenic alignment stages (Figure 4). Once optimized, the point detector was moved out of the way and the diffracted beam was allowed to pass through an objective lens in order to form an image.

## 2.2 DFXM Setup

A DFXM experiment was conducted on a nominally single-crystal sample of $NaMnO_2$ with *in-situ* cooling using a state-of-the-art cryostat, discussed below. The experiment was carried out at 6-ID-C experimental station of the Advanced Photon Source (APS) at Argonne National Laboratory.

A monochromatic X-ray beam from a Si (111) double-crystal monochromator was passed through a pair of entrance slits forming a 300 μm x 300 μm square to uniformly illuminate the sample. Two types of X-ray objective lenses were employed in this work. A beryllium-based objective consisted of a series of 2D parabolic lenses, in groups of 8, or 16, to create a variable compound refractive lens (CRL), which offered a 'zooming' capability. For the low temperature experiment, with a large field of view (FoV), a group of 24 individual lenses were combined to achieve a focal length of 320 mm (10 keV) and a sample to objective distance ($d_{obj}$) of 368.5 mm, resulting in an X-ray magnification of 7.6x at the image plane ($d_{img}$), 2,800 mm away from the sample. The CRL had an effective aperture of 300 μm. To achieve a deeper probe into the bulk at a higher X-ray energy (13 keV), polymer-based lenses with a short focal length (100 mm) [18, 19] were used for the room temperature survey (Figure 3). See Refs. 19-21, for details of CRL 'zoom'-lens design and characterization using Talbot Interferometry.

The far-field detector consisted of an anti-reflection coated LuAG:Ce scintillator (0.02 mm thick and 10 mm diameter) that converted the magnified X-ray beam to visible light, followed by a 45° reflective mirror, a high resolution 5x optical objective and Andor's Zyla 5.5-megapixel sCMOS camera, respectively. The far-field detector is placed at the image plane to capture data in the form of 16-bit, 2560x2160 pixel tiffs. The optical magnification combined with the X-ray magnification resulted in a 38x total magnification using the Be CRL for the temperature dependent measurements, and an 86x total magnification using the polymeric CRL for the room temperature survey. The native pixel size of the sCMOS is 6.5x6.5 μm², so the diffraction limited resolution for these experiments were 170 nm and 76 nm corresponding to 38x and 86x magnifications respectively.

To enable low-temperature *in-situ* DFXM measurements, a custom-modified commercial low-vibration closed-cycle helium refrigerator (Montana

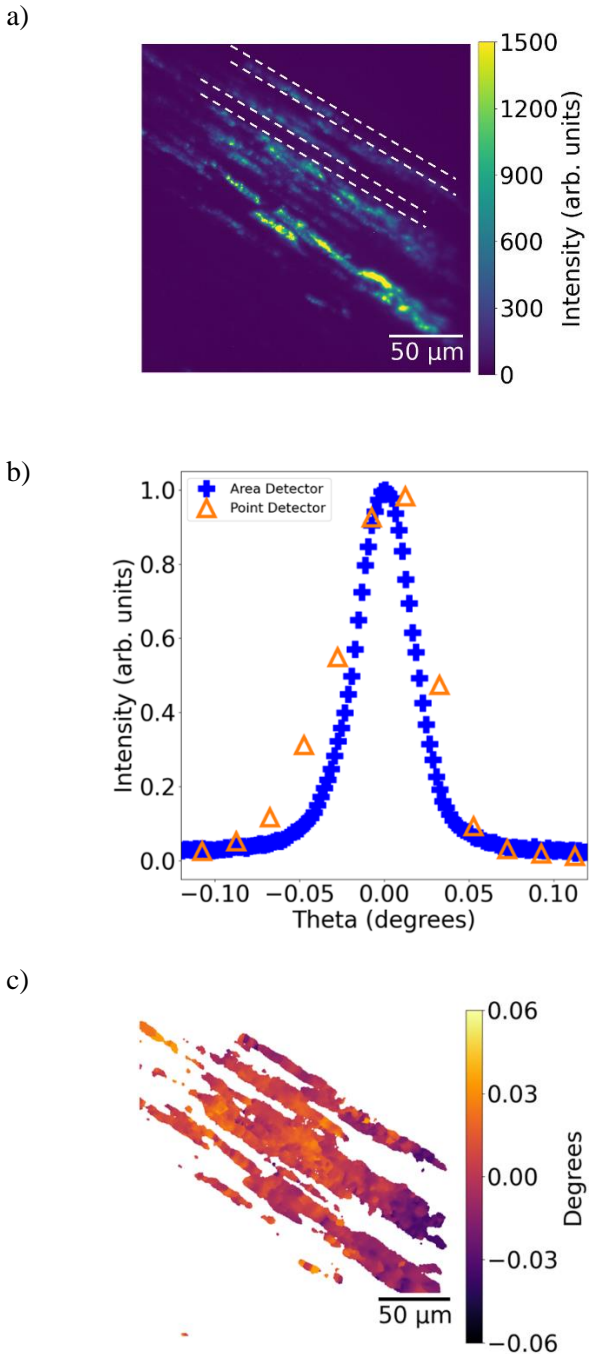

Figure 4: A typical intensity image (a) and corresponding mosaic map (c) captured at room temperature. White dotted lines in (a) guide the eye to defect boundaries. Intensity measurements through the rocking curve scans were collected with a point detector and integrated over the area detector images (b).

Instruments, model s100) was employed. This cryostat provided several features: (i) a set of optical viewports for a complementary VMU, mentioned above, focused on the sample's region of interest (ROI); (ii) 12.5 mm tall beryllium windows on either side to allow for the passage of X-rays at angles of incidence up to 55 degrees and scattering angles up to 110 degrees; (iii) low-vibration (on the order of tens of nanometers) platform mounting; (iv) a base temperature of <3.5 K; and (v) a sample chamber sufficiently large to house a series of nano-positioning stages and goniometer. The cryostat was mounted on top of a diffractometer, using a vibration-damped optical breadboard, which is capable of sample translations in three independent directions and a rotation around the vertical axis ('rock', denoted here as $\theta$) with an orthogonal circle segment ('roll', denoted here as $\chi$), providing a horizontal scattering geometry (Figure 1). The rotational resolution of the diffractometer is 0.0001° with a sample translation resolution, aided by optical encoders, of approximately 100 nm.

## 3. Results & Discussion

DFXM images were collected at each $\theta$ as it was stepped through the (2,0,-2) Bragg peak in 0.002° increments to obtain the material's mosaic spread, known as rocking curve imaging (RCI). RCI is carried out before microscopy to assess the full range of mosaic spread and to ensure that chosen region is a good representation of the entire sample. In Figure 4 we present a representative DFXM image (Fig. 4a), associated rocking curve (Fig 4b), and a derived mosaic map (Fig 4c) from a room-temperature RCI, collected using a Be-CRL with an exposure time of 10 seconds per image. The intensity image Fig. 4a was generated by plotting each pixel's maximum intensity over all frames collected at a given location. Clear contrasts arising from slanted linear twin boundaries, mutually separated by a range of distances, are observed. Furthermore, non-uniform features are visible along the 'ridges' of linear domains. A Bragg-peak line profile obtained by a point detector shows a total mosaic spread of ~0.1 degrees. By integrating each image, a near-identical line profile was obtained. The process of generating a mosaic map, as shown in Fig. 4c, involves mapping material points to individual pixels and comparing their intensity values over variations in sample orientation. As the sample is rocked, crystal domains with varied reciprocal lattice vectors are rotated through the Bragg condition, causing different pixels to illuminate. By finding the center $\theta$ value of each pixel's intensity peak, local regions can be labeled based on their deviation from the principal crystallographic peak [22]. The mosaic map in

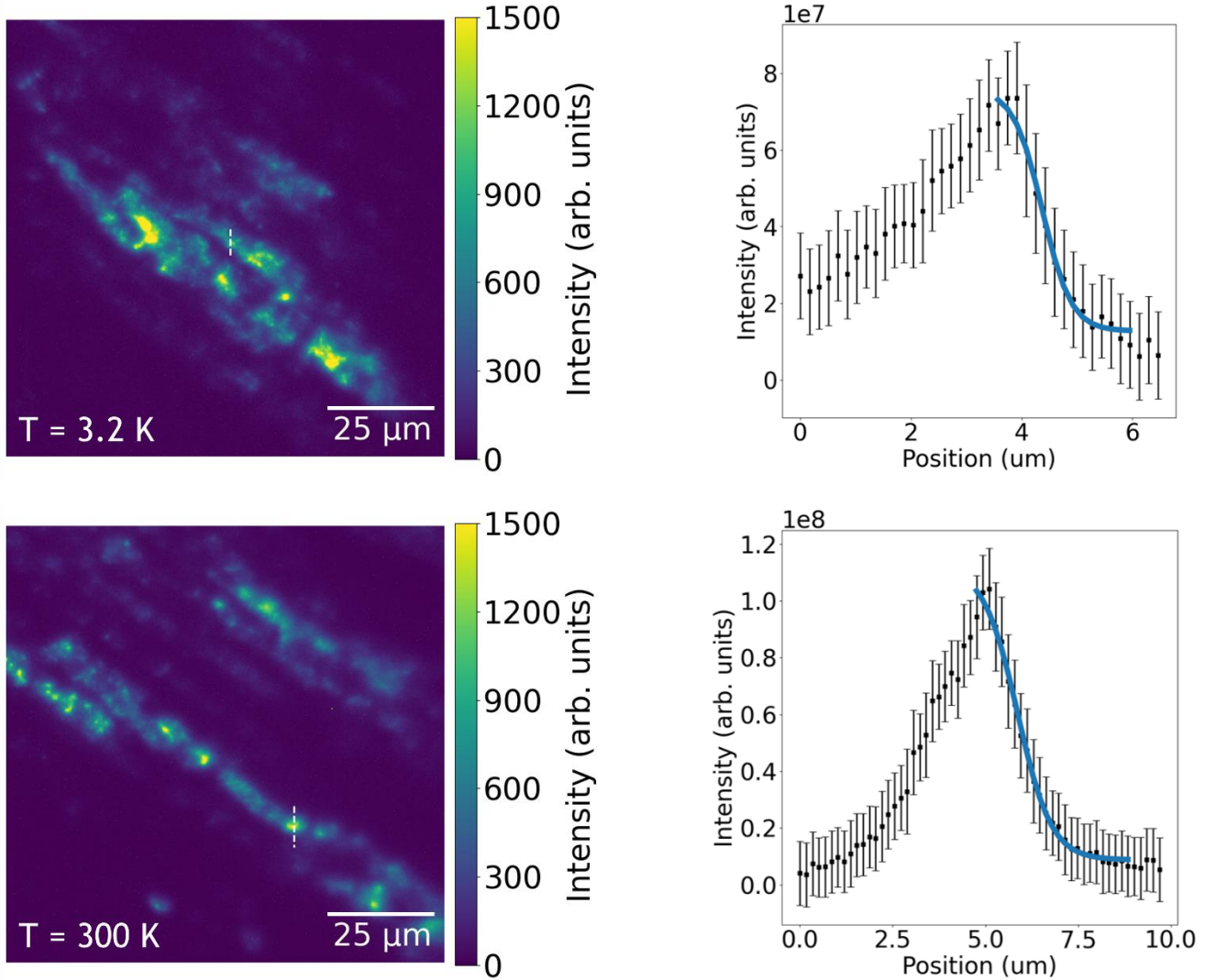

Figure 5: Diffraction contrast images (left) taken at base temperature (top) and room temperature (bottom) using the Be CRL objective lens. White dotted lines indicate where the line profiles (right) were taken for each image. Logistical fits to the line profiles are shown and the FWHM of their derivatives was calculated to estimate the smallest resolvable feature size. In this case, images at both temperatures displayed a feature resolution of ~1 μm or ~6 pixels.

Figure 4c shows a gradual distribution of local orientation with an overall spread consistent with the Bragg-peak profile discussed above. These measurements, performed over spots shown in Figure 3, indicate that the sample does not have drastic changes in its mosaic distributions across its millimeter length scale. Therefore, it is expected that the results of the low-temperature experiment detailed below, performed on an arbitrary spot on the sample, are representative of properties of $NaMnO_2$ in general.

The triclinic nanodomains are reported to occur below the magnetic ordering temperature of 45 K, with an additional magnetic transition existing at 22 K [5,15]. In this proof-of-principle work, we focused our efforts primarily on the structural transition at 45 K, for which magneto-structural effects are likely to be less subtle.

For the temperature dependent experiment, we used a Be-CRL with a 300 μm diameter field of view, while still resolving micron-scale features (see Figure 5) at all temperatures (3.2-300 K). The same region of the sample was identified through temperature changes using optical images, via the VMU mentioned above. A shallow depth-of-field (10 μm) combined with the high-resolution (345 nm per pixel) of the optical microscope, a similar scale as the DFXM images (see Figure 5), provides confidence that sample position was maintained during temperature-dependent measurements.

A series of RCI was then carried out at various temperatures from 3.2 K through low-temperature transitions to detect any local changes in strain and/or orientation due to a purported structural transition at 45 K. These raw images were used to generate 'mosaic' maps, or maps of the spread in orientation of the sample at various temperatures. Mosaic maps generated from RCI, done before and after the transition temperature, are shown in Figure 6. There is a notable (-θ) shift in the orientation of several regions as the sample is heated from 9 K to 52 K, indicating that the transition impacts the mosaic spread of the material at a local level, possibly causing an average broadening of the (2,0,-2) peak.

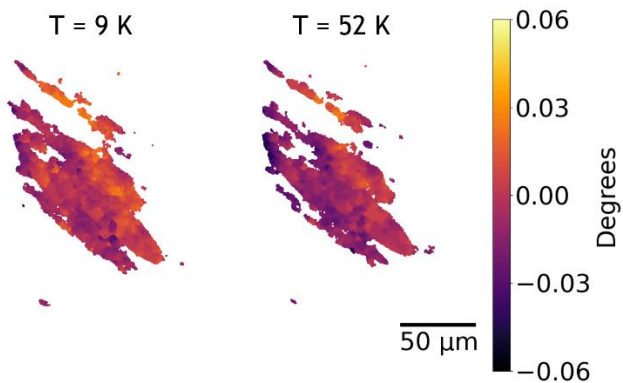

Figure 6: Evolution of mosaic maps through sample heating, generated from DFXM images taken during rocking curve scans.

To make our observations more quantitative, temperature-dependent measurements, focused on this transition, were taken by holding θ fixed at the peak value and capturing images at a constant rate as the sample was slowly heated from 10 K to 52 K. The total intensity of the images was calculated and plotted against temperature in Figure 7. A logistic curve fit to the data is overlaid and shows a significant jump (20% increase) in the intensity of the total (2,0,-2) peak, centered on 44 K, spanning over a temperature range of ~10 K. It is intriguing that while certain regions are significantly affected through the transition, the rest of our FoV remained quiescent. The regions outlined by boxes undergo an overall 55% and 25% increase through the transition for the first and second ROIs respectively.

The magnified images show subtle changes in various features as the sample was heated through the transition, but it remains difficult to say whether these changes are direct observations of nanoscale triclinic domains undergoing a transition to the primary monoclinic phase. These effects may also be an example of a magnetoelastic coupling, which results in structural strain or rotation due to the magnetic transition. Additionally, the $Mn_3O_4$ intergrowth, mentioned above, exhibits a weak ferrimagnetic transition at 43 K [23]. We suspect that the intergrowth forms in stacking fault regions or twin boundaries. Thus, the transition from para- to ferrimagnetism could result in magneto-striction of the $NaMnO_2$ lattice, causing certain regions to 'fall off' the primary Bragg-peak. While previous studies of this transition have measured powders, the spatially resolved images collected using DFXM on a single crystal sample give the first indication of local heterogeneities playing a significant role in the magneto-structural transition at low temperatures.

## 4. Conclusions

DFXM is a versatile and powerful technique to study spatially resolved 'mesoscale' phenomena of functional, structural, and quantum materials, complementing observations of 'average' properties using X-ray diffraction in routine use. The experiment detailed here is the first low-temperature DFXM study reaching liquid-helium temperature on a highly heterogeneous single-crystal sample at a third-generation light source, which enables spatially resolved studies of twins, dislocations, charge-order or magnetic domains through a wide array of thermal phase transitions. The outlined experimental procedure lays the foundation for the characterization and evaluation of other functional and structural materials at low temperatures.

The preliminary low-temperature investigation of $NaMnO_2$ provided valuable insight into its possible structural transition at 45 K, which has been a challenge to unambiguously detect. We speculate that a 'phase transformation' might be taking place in privileged regions within an intertwined hierarchy of three-dimensional structural and phase heterogeneities spanning a wide range of length scales. This is consistent with significant local changes observed herein. More comprehensive DFXM studies of spatially resolved strain as a function of temperature may indicate a mechanism for intensity step through transition. Furthermore, complementary imaging methods [24, 25] such as high-energy diffraction microscopy performed on the specimen, may isolate and identify these active regions susceptible to this transition.

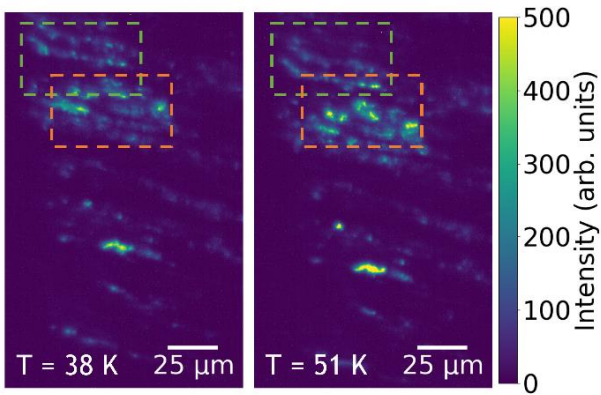

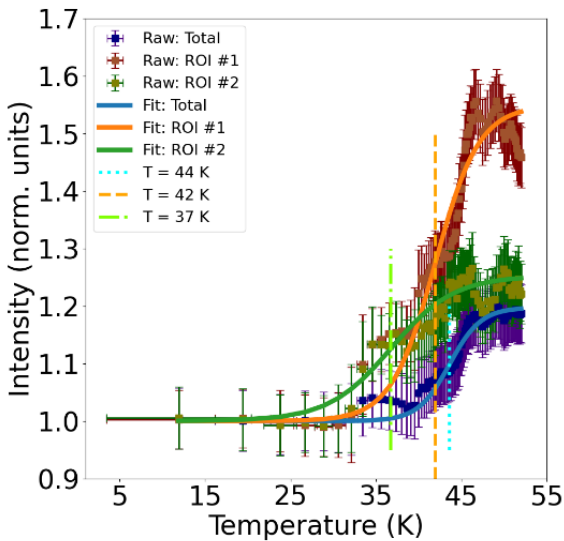

Figure 7: DFXM images taken while holding the sample orientation constant and at the maximum intensity for two temperatures, one before (a) and one after (b) the supposed transition temperature of 45 K. Orange and green dashed boxes indicate ROIs that were selected for comparison. Intensity values are plotted in (c) against temperature for the primary (2,0,-2) peak, computed for the total image and for the ROIs displayed in blue, orange and green respectively. The data includes error bars for both temperature and intensity fluctuations. The transition resulted in a 20%, 55% and 25% increase in the overall and regional intensities, and inflection points at 44, 42 and 37 K respectively.

## Data Availability

The raw/processed data required to reproduce these findings cannot be shared at this time as the data also forms part of an ongoing study.

## Acknowledgments

The research reported here was supported by the National Science Foundation (NSF) Materials Research Science and Engineering Center (MRSEC) at UC Santa Barbara (NSF DMR 1720256) through IRG-1. We acknowledge the use of shared facilities of the NSF MRSEC at UC Santa Barbara [DMR 1720256] and the Center for Scientific Computing, supported by the California Nano Systems Institute, the NSF MRSEC (DMR 1720256) and NSF CNS 1725797. Travel support was provided via the U.S. Department of Energy, Office of Science, Office of Science Graduate Student Research (SCGSR) program. The SCGSR program is administered by the Oak Ridge Institute for Science and Education (ORISE) for the DOE. ORISE is managed by ORAU under contract number DE-SC0014664. All opinions expressed in this paper are the authors' and do not necessarily reflect the policies and views of DOE, ORAU, or ORISE. This research used resources of the Advanced Photon Source, a U.S. Department of Energy (DOE) Office of Science User Facility at Argonne National Laboratory and is based on research supported by the U.S. DOE Office of Science-Basic Energy Sciences, under Contract No. DE-AC02-06CH11357. The identification of any commercial product or trade name does not imply endorsement or recommendation by the National Institute of Standards and Technology. The authors would like to thank the Karlsruhe Nano Micro Facility (KNMF) for the fabrication of the polymer X-ray optics.

## References

[1] - Crabtree, G., Sarrao, J., Alivisatos, P., Barletta, W., Bates, F., Brown, G., ... & Tranquada, J. (2012). *From quanta to the continuum: opportunities for mesoscale science*. USDOE Office of Science (SC)(United States).

[2] - Simons, H., Jakobsen, A. C., Ahl, S. R., Detlefs, C., & Poulsen, H. F. (2016). Multiscale 3D characterization with dark-field x-ray microscopy. *Mrs Bulletin*, *41*(6), 454-459.

[3] - Yildirim, C., Cook, P., Detlefs, C., Simons, H., & Poulsen, H. F. (2020). Probing nanoscale structure and strain by dark-field x-ray microscopy. *MRS Bulletin*, *45*(4), 277-282.

[4] - Ko, W., Gai, Z., Puretzky, A. A., Liang, L., Berlijn, T., Hachtel, J. A., ... & Li, A. P. (2022). Understanding Heterogeneities in Quantum Materials. *Advanced Materials*, 2106909.

[5] - Dally, R. L., Heng, A. J., Keselman, A., Bordelon, M. M., Stone, M. B., Balents, L., & Wilson, S. D. (2020). Three-

magnon bound state in the quasi-one-dimensional antiferromagnet α-NaMnO 2. *Physical Review Letters*, *124*(19), 197203.

[6] - Jakobsen, A. C., Simons, H., Ludwig, W., Yildirim, C., Leemreize, H., Porz, L., ... & Poulsen, H. F. (2019). Mapping of individual dislocations with dark-field X-ray microscopy. *Journal of Applied Crystallography*, *52*(1), 122-132.

[7] - Dresselhaus-Marais, L. E., Winther, G., Howard, M., Gonzalez, A., Breckling, S. R., Yildirim, C., ... & Poulsen, H. F. (2021). In situ visualization of long-range defect interactions at the edge of melting. *Science Advances*, *7*(29), eabe8311.

[8] - Simons, H., Jakobsen, A. C., Ahl, S. R., Poulsen, H. F., Pantleon, W., Chu, Y. H., ... & Valanoor, N. (2019). Nondestructive mapping of long-range dislocation strain fields in an epitaxial complex metal oxide. *Nano letters*, *19*(3), 1445-1450.

[9] - Bucsek, A., Seiner, H., Simons, H., Yildirim, C., Cook, P., Chumlyakov, Y., ... & Stebner, A. P. (2019). Sub-surface measurements of the austenite microstructure in response to martensitic phase transformation. *Acta Materialia*, *179*, 273-286.

[10] - Abakumov, A. M., Tsirlin, A. A., Bakaimi, I., Van Tendeloo, G., & Lappas, A. (2014). Multiple twinning as a structure directing mechanism in layered rock-salt-type oxides: NaMnO2 polymorphism, redox potentials, and magnetism. *Chemistry of Materials*, *26*(10), 3306-3315.

[11] - Xu, G. L., Liu, X., Zhou, X., Zhao, C., Hwang, I., Daali, A., ... & Amine, K. (2022). Native lattice strain induced structural earthquake in sodium layered oxide cathodes. *Nature communications*, *13*(1), 1-12.

[12] - Zorko, A., Adamopoulos, O., Komelj, M., Arčon, D., & Lappas, A. (2014). Frustration-induced nanometre-scale inhomogeneity in a triangular antiferromagnet. Nature Communications, 5(1), 1-10.

[13] - Dally, R., Clément, R. J., Chisnell, R., Taylor, S., Butala, M., Doan-Nguyen, V., ... & Wilson, S. D. (2017). Floating zone growth of α-Na0. 90MnO2 single crystals. *Journal of Crystal Growth*, *459*, 203-208.

[14] - Dally, R. L., Zhao, Y., Xu, Z., Chisnell, R., Stone, M. B., Lynn, J. W., ... & Wilson, S. D. (2018). Amplitude mode in the planar triangular antiferromagnet Na0. 9MnO2. *Nature communications*, *9*(1), 1-8.

[15] - Dally, R. L., Chisnell, R., Harriger, L., Liu, Y., Lynn, J. W., & Wilson, S. D. (2018). Thermal evolution of quasi-one-dimensional spin correlations within the anisotropic triangular lattice of α− NaMnO 2. *Physical Review B*, *98*(14), 144444.

[16] - Preibisch, S., Saalfeld, S., & Tomancak, P. (2009). Globally optimal stitching of tiled 3D microscopic image acquisitions. *Bioinformatics*, *25*(11), 1463-1465.

[17] - Schindelin, J., Arganda-Carreras, I., Frise, E., Kaynig, V., Longair, M., Pietzsch, T., … Cardona, A. (2012). Fiji:an open-source platform for biological-image analysis. *Nature Methods*, *9*(7), 676–682. doi:10.1038/nmeth.2019

[18] - Opolka, A., Müller, D., Fella, C., Balles, A., Mohr, J., & Last, A. (2021). Multi-Lens Array Full-Field X-ray Microscopy. *Applied Sciences*, *11*(16), 7234.

[19] - Qiao, Z., Shi, X., Kenesei, P., Last, A., Assoufid, L., & Islam, Z. (2020). A large field-of-view high-resolution hard x-ray microscope using polymer optics. *Review of Scientific Instruments*, *91*(11), 113703.

[20] - Xianbo Shi, Walan Grizolli, Deming Shu, Luca Rebuffi, Zahirul Islam, and Lahsen Assoufid. "*High-speed characterization of refractive lenses with single-grating interferometry*." Proc. of SPIE 11109 (2019); doi: 10.1117/12.2529886

[21] - D. Shu, Z. Islam, J. Anton, S. Kearney, X. Shi, W. Grizolli, P. Kenesei, S. Shastri, and L. Assoufid, "*Mechanical design of a new precision alignment apparatus for compact x-ray compound refractive lens manipulator*," 10[th] Mechanical Eng. Design of Synchrotron Radiation Equipment and Instrumentation, JACoW 168-172 (2018). DOI: 10.18429/JACoW-MEDSI2018-WEOPMA04

[22] - Ahl, S. R. (2018). Elements of a Method for Multiscale Characterization of Recrystallization in Deformed Metals.

[23] - Boucher, B., Buhl, R., & Perrin, M. (1971). Magnetic structure of Mn3O4 by neutron diffraction. *Journal of Applied Physics*, *42*(4), 1615-1617.

[24] - Pokharel, R. (2018). Overview of High-Energy X-Ray Diffraction Microscopy (HEDM) for Mesoscale Material Characterization in Three-Dimensions. In: Lookman, T., Eidenbenz, S., Alexander, F., Barnes, C. (eds) Materials Discovery and Design. Springer Series in Materials Science, vol 280. Springer, Cham. https://doi.org/10.1007/978-3-319-99465-9_7

[25] - Liu, H., Zhang, Y., Wilkin, M., Park, J. S., Kenesei, P., Rollett, A. D., & Suter, R. M. (2022, July). 3D In-situ Stop Action Study of Recrystallization in Additively Manufactured 316L Stainless Steel: Reconstruction Optimization and Observations. In *IOP Conference Series: Materials Science and Engineering* (Vol. 1249, No. 1, p. 012054). IOP Publishing.